\documentstyle[12pt]{article}
\def\be{\begin{equation}}
\def\ee{\end{equation}}
\def\ba{\begin{eqnarray}}
\def\ea{\end{eqnarray}}
\def\bq{\begin{quote}}
\def\eq{\end{quote}}
\def\PL{{ \it Phys. Lett.} }
\def\PRL{{\it Phys. Rev. Lett.} }
\def\NP{{\it Nucl. Phys.} }
\def\PR{{\it Phys. Rev.} }

\begin{document}

\renewcommand{\theequation}{\arabic{section}.\arabic{equation}}
\newcommand{\dftwo}{(\nabla\phi)^2}
\newcommand{\dstwo}{(\nabla\sigma)^2}
\newcommand{\dffour}{(\nabla\phi)^4}
\newcommand{\riemtwo}{R_{\mu\nu\sigma\rho}R^{\mu\nu\sigma\rho}}
\newcommand{\rhh}{R^{\mu\nu\sigma\rho}
H_{\mu\nu\alpha}H_{\sigma\rho}{}^{\alpha}}
\newcommand{\hfour}{H_{\mu\nu\lambda}H^{\nu}{}_{\rho\alpha}
H^{\rho\sigma\lambda}H_{\sigma}{}^{\mu\alpha}}
\newcommand{\htwohtwo}{H_{\mu\rho\lambda}
H_{\nu}{}^{\rho\lambda}H^{\mu\sigma\alpha}H^{\nu}{}_{\sigma\alpha}}
\newcommand{\dP}{{\dot\phi}}\
\newcommand{\tr}{{\rm Tr}}
\newcommand{\si}{{\sigma}}
\newcommand{\ddt}{\frac{\rm d}{{\rm d} t}}
\newcommand{\pa}{\nabla}
\newcommand{\lz}{\lambda_0}
\newcommand{\ets}{{e^{2\sigma}}}
\newcommand{\emts}{{e^{-2\sigma}}}
\newcommand{\aplz}{\alpha'\lambda_0}

\thispagestyle{empty}
\begin{flushright}
CERN-TH/97-113\\
WATPHYS-THY-96/18\\
%hep-th/xxx
\end{flushright}
\vspace*{1cm}
\begin{center}
{\Large \bf  Duality Beyond the First Loop}
 \\
\vspace*{1cm}
Nemanja Kaloper\footnote{E-mail: kaloper@hepvs6.physics.mcgill.ca}\\
\vspace*{0.2cm}
{\it Department of Physics, University of Waterloo}\\
{\it Waterloo, ON N2L 3G1, Canada}\\
\vspace*{0.4cm}
Krzysztof A. Meissner
\footnote{Permanent address: Institute for Theoretical Physics,
Ho{\.z}a 69, 00-689 Warszawa, Poland.\\
\indent ~~E-mail: Krzysztof.Meissner@cern.ch}\\
\vspace*{0.2cm}
{\it CERN, 1211 Geneva 23, Switzerland}\\
\vspace{2cm}
ABSTRACT
\end{center}
In this article we give a calculation of the
two-loop $\si$-model corrections to the $T$-duality map 
in string theory. We use the effective action 
approach, and analyze two-loop corrections
in a specific subtraction scheme.
Focusing on backgrounds which have a single Abelian
isometry, we find the explicit form for the $O(\alpha')$ modifications 
of the lowest order duality transformations. 
Rather surprisingly, the manifest two-loop
duality depends crucially on the torsion field.
In contrast to the dilaton and metric fields,
which are merely passive spectators, the torsion 
plays a more active role, because of the anomalous couplings
to the gauge fields that arise via dimensional reduction.
Our results support the interpretation of $T$-duality
as an expansion in the inverse string tension $\alpha'$, and 
its order-by-order realization as 
a manifest symmetry of the full string theory. 
\vfill
\begin{flushleft}
CERN-TH/97-113\\
May 1997
\end{flushleft}
\setcounter{page}{0}
\setcounter{footnote}{0}

\section{Introduction}
Non-compact continuous symmetries
different from the trivial 
diffeomorphism group are very rare in
gravitational systems. An example 
was discovered by Ehlers for 
the case of the four-dimensional 
vacuum Einstein gravity with a single 
Killing vector. These results were 
later extended by Geroch \cite{geroch}
to the solutions with two Killing 
isometries, where the symmetry is
enhanced to an infinite-dimensional
Kac-Moody algebra. In all of 
these cases, however, non-compact symmetries 
arise as the hidden symmetries
of the equations of motion. The dynamics 
of gravitational systems admits
hidden integrals of motion, which generate the 
non-compact symmetry by acting on 
different momentum modes of the gravitational 
field. In this sense, the Geroch
group and its clones are intrinsic to the 
theory of gravity - they only 
permute the gravitational degrees of freedom, 
while leaving the overall
theory unaffected.

On the other hand, the gravitational 
multiplet in string theory
has a considerably richer structure than 
in pure Einstein gravity. In
addition to the graviton, it always contains 
a scalar (dilaton) and an 
antisymmetric tensor (Kalb-Ramond field, also 
often referred to as 
the torsion). Depending on the type of 
string theory at hand, there
may be additional modes, describing the 
gauge sector of the theory.
In this work we will focus on the 
model-independent sector of the massless
modes, leaving the issue of the other modes 
for later. The low-energy
dynamics of these modes is given by 
an effective action (identical for all consistent 
string truncations to the 
model-independent modes in the lowest order),
which is covariant, and incorporates 
the dilaton and torsion effects via 
nonminimal couplings. The resulting
theory can be thought of as 
gravity with corrections, coming from the
stringy degrees of freedom. 
As this interpretation suggests, 
the symmetries of the Ehlers and 
Geroch type generalize straightforwardly 
to the case of the effective actions 
in string theory \cite{bakas}. The enhanced 
complexity of the phase space of string
theory however may be taken 
to indicate that string dynamics 
could be even more symmetric than its
pointlike counterpart. Indeed, in string 
theory there appears yet another class 
of symmetries. These symmetries have by 
now become widely known under the
general name of duality, and have 
emerged as an indispensable tool
in the quest for a better 
understanding of string theory (a.k.a. 
``theory formerly known as strings"). 
The impact of these symmetries
on the dynamics of string theory is 
far more profound than the impact of the
hidden dynamical symmetries in 
General Relativity, because dualities 
provide the natural maps between 
seemingly different string theories, and
not only different solutions in the 
phase space of a single theory.
Specifically, the greatest achievement 
of dualities to date has been the 
development of the ``web of dualities" (see for 
example \cite{witten}), which is the basis of the 
uniqueness proof of string theory, whereby 
all consistent string constructions 
have been recognized as the facets  
of a single fundamental theory, labeled the M-theory. 

A duality symmetry of interest 
to us here will be the so-called
$T$-duality, or string scale-factor 
duality. At the level of the 
world-sheet $\sigma$-model, this 
symmetry has been identified in
\cite{buscher} as a simple Hodge-type 
duality of a cyclic target coordinate,
generated by its flows. The full 
effect of such transformations on a generic set
of target-space 
degrees of freedom has been exhibited
in \cite{mv12} (in the cases with 
fewer degrees of freedom, 
a discrete symmetry of the action 
was found in \cite{ven} and \cite{ts91}).
In these articles it has been 
shown that when fields depend on
only a single coordinate in 
an arbitrary number of space-time dimensions,
the lowest-order Lagrangian exhibits 
continuous, global $O(d,d,R)$ symmetry. 
The symmetry has subsequently been
extended to include matter \cite{gv} 
and/or gauge fields \cite{hassen,mahsch} 
(we should also mention an early 
analysis in \cite{duff}, which however
has ignored the dilaton field) and 
has been identified in all string constructions. 
There it has also been realized that the continuous
$O(d,d,R)$ symmetry is broken to its discrete $O(d,d,Z)$
subgroup by instanton-like effects.
Finally, it has been discovered
that the discrete ${\bf Z_2}$ subgroup
of this group takes a distinguished place
in the above-mentioned ``web of dualities" \cite{witten}.
In this context,  $T$-duality is not a map in the 
phase space of a single theory, but instead a map 
between different theories - an example of 
the action of such a duality are
type IIA and
type IIB superstring theories compactified on 
$K3$ down to six dimensions.

Most of the explicit investigations and 
applications of $T$-duality were conducted
at the one-loop level of the 
effective approach to string theory, after 
truncating the action down to 
the terms of second order in derivatives. 
Yet, it has been indicated, and in some special
situations proven, that this symmetry
is exact order-by-order 
in perturbation theory. In \cite{mv12}
an argument has been given that the symmetry
ought to be present to all 
orders in $\alpha'$ in the $\sigma$-model
expansion; another argument 
has been offered in the second reference of \cite{hassen}. 
Further, in \cite{ts91} 
it has been argued that in the absence
of torsion there should be 
corrections to fields in the next
order of $\alpha'$ to ensure 
that the $\beta$-functions vanish, and in
\cite{ts91,panvel} specific 
examples to this effect have been given.
A more general study of such 
solutions has been undertaken in \cite{km},
where the first explicit proof of $T$-duality to two
loops has been given. There, the unique $O(\alpha')$ 
effective action consistent with unitarity and
$T$-duality invariance has been derived, 
and the connection between $T$-duality 
and the ``off-shell'' $\beta$ function approach has been
pointed out (at one loop such a connection was noticed in \cite{haag1}).
Later, other work has been done in the attempt to
understand the corrections to duality away from
conformal points \cite{sigmamod}.

Lastly, it has been shown that the 
lowest-order form of the on-shell $T$-duality
map remains unaffected by higher 
order $\alpha'$ corrections when viewed
as a relation between specific conformal field
theories (CFT's) \cite{ekir}, some dual 
solutions in two dimensions \cite{tsey2d} and
some special supersymmetric solutions 
\cite{exact}. In these cases, the proof relied
on special properties of the 
solutions studied - there either existed an
exact, nonperturbative CFT formulation, 
or the solutions were highly symmetric,
which protected them from acquiring quantum corrections. 
A picture which has emerged from these examples 
is that the $T$-duality map can be expanded as 
a perturbative 
series in the inverse string tension $\alpha'$.

The overwhelming importance 
of dualities then seems to beg the 
question of their validity beyond 
the first loop\footnote{We would like to underline
here that we consider the loop expansion of the
world-sheet $\sigma$-model in the field-theoretic sense -
hence associating $\alpha'$ with the loop counting variable.
An alternative interpretation of these 
corrections, from the target-space
viewpoint, is that they are 
classical, since $\alpha'$ 
measures deviations from the pointlike theory due to the
extended structure of strings.
We are, however, not going to consider 
the genus expansion, i.e. the true 
quantum-theoretic sector of string theory.}. 
Since the inclusion of 
the next-order terms (e.g., 
curvature squared etc.) can be very 
important for the stability 
of solutions (as has been
recently pointed out in the 
case without torsion \cite{gmagv}), 
and perhaps even elucidate 
some of the quantum-mechanical properties
of string theory
\footnote{A good example is the 
compatibility between duality and
anomaly cancellation. To the 
lowest order, this is illustrated by
the explicit appearance of the 
generalized gauge Chern-Simons terms in the
reduced torsion \cite{mahsch}, 
on one hand dictated by the form of the theory
and on the other hand necessitated by duality. 
While one may expect that this
survives quantum corrections, an explicit 
check is always more convincing.}, 
it is of interest to test the influence of 
the $\alpha'$ corrections on the 
$T$-duality maps.

The purpose of this work is to explicitly show 
that $T$-duality remains manifest 
(while acquiring $O(\alpha')$ 
corrections) even when 
the two-loop effects are accounted 
for. The calculation 
we will present below is a natural 
extension of Buscher's \cite{buscher} 
results beyond the first loop, and 
is based on identical assumptions:
we only restrict our backgrounds to 
possess a single isometry, compact or not,
and take into account all the 
resulting reduced zero-mass excitations: the metric, 
dilaton, torsion, and momentum 
and winding gauge fields. We then take an array of two-loop
counterterms, and carry out dimensional reduction on 
the isometry. The ensuing reduced action contains 
several terms which appear to violate the
duality invariance of the action to 
$O(\alpha')$. These terms force us to introduce
the $O(\alpha')$ corrections of the 
$T$-duality map: once properly accounted for, 
$T$-duality reemerges as the 
invariance of the theory to two loops. A helpful
organizing principle, and a 
consistency check, in this calculation 
is  the gauge invariance of 
the reduced theory. Inherited from higher dimensions 
from a subgroup of the full 
diffeomorphism group and the torsion gauge group, 
the reduced gauge invariances 
are anomalous because of the non-invariant split of the
components of the torsion 
potential, which however admits a gauge-invariant field
strength \cite{mahsch}. The Chern-Simons
gauge terms which must be included 
in the torsion field precisely cancel the
gauge dependence of the torsion 
potential. But they are also crucial for the
theory, for without them $T$-duality 
would have been lost. For example, in 
heterotic string theory they are 
partnered up with the vector supermultiplet Yang-Mills
Chern-Simons terms, and only 
in their presence is the generalization of $T$-duality, 
the full $O(d,d+n)$ symmetry, involving
permutations of the gauge 
field sector, manifest. Because of this, we can 
develop a consistency check of our calculations. 
In the heterotic theory the Yang-Mills Chern-Simons 
arise via chiral anomaly, which can be computed with the 
help of the target-space triangle diagrams. These terms
however appear only at the one-loop level of the 
target space theory, and do not receive any further
corrections. If $T$-duality is to 
be valid, similar conditions must be met by
the Kaluza-Klein gauge modes as well. 
Such restrictions on the Kaluza-Klein gauge
corrections will turn out to be
very helpful in determining the precise 
form of the $O(\alpha')$ corrections.
We will see that the form of the anomaly can be 
consistently preserved, and that it does not
develop any new, non-trivial, $O(\alpha')$ corrections.
Further, using this we will see how the 
anomaly transformation rules in fact assist in
maintaining $T$-duality to $O(\alpha')$.

The paper is organized as follows. In the next section 
we review the one-loop results, in order to set the
stage for the realm of $O(\alpha')$ terms. Section 3. is 
devoted to the detailed investigation of the $O(\alpha')$
corrections. Specifically, we will carry out the 
reduction of the effective action on a single cyclic
coordinate, and isolate the terms which break the
lowest order duality symmetry. Using these results,
we will derive the explicit form of the $O(\alpha')$
corrections to the $T$-duality map in section 4.
This will in effect prove that $T$-duality is valid 
to two loops, on any class of backgrounds which 
admit the lowest order symmetry. Finally, in the
last section we will present our conclusions, and
indicate possibilities for future extensions of our
results.

\section{One-loop $T$-duality}

\setcounter{equation}{0}
We begin here by reviewing the derivation of the
one-loop $T$-duality transformations to set the 
notation. In doing so, we are following 
the computation in \cite{mahsch}.
The lowest-order term in the 
effective action of any string theory
truncated to only the model-independent 
zero mass modes is (throughout this paper we use the
string frame with $e^{-2\phi}$ out front, since the symmetry is
most simply realized there)
\be
\Gamma^{(0)} = \int d^{d+1}x \sqrt{\bar g}~ e^{-2 \bar \phi}
\left\{\bar R(\bar g)+4 (\bar \nabla \bar \phi)^2 
-\frac{1}{12}\bar H^2\right\}.
\label{actnor}
\ee
Our convention for the signature of the 
metric is $(-,+,\ldots,+)$, the Riemann
curvature is 
$\bar R^{\mu}{}_{\nu\rho\sigma}=\partial_{\rho}
\bar \Gamma^{\mu}{}_{\nu\sigma}-\ldots$,
and the torsion field strength is the 
antisymmetric derivative of the torsion potential:
$\bar H_{\mu\nu\rho}=\nabla_{\mu}\bar B_{\nu\rho}+{\rm cyclic}$. 
The overbar denotes the quantities in the original, $d+1$-dimensional,
frame, before we carry out Kaluza-Klein reduction.
Note that the definition of the
torsion field strength $\bar H$ encodes the torsion 
potential gauge invariance: if we shift 
the $\bar B$-field according to 
$\bar B \rightarrow \bar B + d \Lambda$, where $\Lambda$ is an arbitrary
one-form, the theory remains unchanged. 
As a consequence, the Bianchi
identity for $\bar H$ takes a very simple form: $d\bar H=0$. 

In the presence of a single 
Killing isometry, we can carry out 
the Kaluza-Klein dimensional reduction down to
$d$ dimensions. The notion which best charts this 
dimensional descent is gauge invariance. 
It is not hard to see that the ``off-diagonal" 
terms in the metric and the torsion potential
(those terms mixing the cyclic with non-cyclic 
indices) must appear as gauge fields in lower
dimensions. The reason for this is that 
there are Abelian gauge invariances associated with
them. The translations 
of the form $y \rightarrow y'=y-\omega(x)$ 
where $y$ is the cyclic coordinate and $x$'s 
denote the non-cyclic ones, and the special
gauge transformations of the two-form $B$ 
with $\Lambda = \lambda_y(x) dy$ leave the theory 
invariant, but after integrating out the 
cyclic coordinate $y$, they must appear 
as the matter sector symmetries, 
acting on the vectors which emerge from 
the metric and torsion. Hence the most natural 
way to carry out the reduction is to 
ensure that these symmetries are manifest at every step 
of the calculation - this would guarantee 
that the final result is symmetric as well.
Thus the reduced action will feature two additional 
gauge fields coupled
to the graviton, axion and dilaton.
Moreover, there will also appear an additional
scalar field, which is the breathing mode
of the cyclic coordinate. 
In a more general setting with several commuting
Killing isometries, there would 
also appear scalar modes from the the
reduction of the two-form torsion 
potential, living in the submanifold spanned
by the isometries. In our case, 
they are absent by virtue of their antisymmetry,
and the fact that we reduce on a single
cyclic coordinate.

We can now present the steps of the reduction procedure.
As is well known, gauge invariance is enforced as
follows. One chooses the reduced metric and dilaton, 
$g_{\mu\nu}$ and $\phi$ respectively, according to 
\begin{eqnarray}
\label{dans}
d{\bar{s}}^2 = {g}_{\mu\nu} dx^{\mu} dx^{\nu} 
+ \exp(2 \sigma)
(dy + V_{\mu} dx^{\mu})^2 ~~~~~~~ 
\phi = \bar \phi - \frac{1}{2} \sigma 
\end{eqnarray}
All the lowering and raising of the 
indices in the rest of this
work will be done with respect to the 
reduced metric $g_{\mu\nu}$.
The vector field $V_{\mu}$ is the 
standard Kaluza-Klein gauge field, which 
couples to the momentum modes of the theory.
The reduction of the axion 
field has to be done with more care because of 
the anomaly which appears in it. 
Namely, if we look at the naive decomposition
of the two-form 
$\bar B = (1/2) \bar B_{\mu\nu} dx^{\mu} \wedge dx^{\nu} 
+ W_{\mu} dx^{\mu} \wedge dy$ 
(here $W_{\mu} = \bar B_{\mu y}$ is the other gauge field,
arising from the ``off-diagonal" 
components of the torsion, and which couples to the winding modes) 
and attempt to interpret the space-time
components $\bar B_{\mu\nu}$ as the reduced 
torsion, we would run into problems
with gauge invariance. Whereas they constitute a
gauge-invariant tensor with respect 
to the gauge transformations induced
by any $\Lambda = \Lambda_{\mu} dx^{\mu}$ 
and $\lambda = \lambda_y dy$, 
they are not invariant under the 
translations along $y$, because of the presence of 
$dy$ in the decomposition of the
original, gauge-invariant, torsion 
two-form $\bar B$. We can see
that when $y \rightarrow y'= y - \omega(x)$ 
and $V_{\mu} \rightarrow
V'_{\mu} + \partial_{\mu} \omega$ 
(s.t. the cyclic einbein $E=dy + V_{\mu} dx^{\mu}$
is gauge-invariant), the space-time 
components transform according to
$\bar B_{\mu\nu} \rightarrow \bar B'_{\mu\nu} = 
\bar B_{\mu\nu} + W_{\mu} \partial_{\nu} \omega -
W_{\nu} \partial_{\mu} \omega$. One may be 
tempted to alter the reduced torsion 
potential in analogy 
with the gauge-invariant reduced metric (\ref{dans}).
Indeed, it is straightforward to verify that 
$\hat B_{\mu\nu} = \bar B_{\mu\nu} - 
(W_{\mu} V_{\nu} - W_{\nu} V_{\mu})$ does not
change by the local translations 
along $y$. However, this field is not
invariant under the reduced gauge 
transformations $\lambda_y$.
When $W_{\mu} \rightarrow W'_{\mu} = 
W_{\mu} + \partial_{\mu} \lambda_y$, 
$\hat B_{\mu\nu} \rightarrow 
\hat B'_{\mu\nu} = \hat B_{\mu\nu} -
(\partial_{\mu} \lambda_y  V_{\nu} - 
\partial_{\nu} \lambda_y V_{\mu})$.
As a result, we see that the two gauge
symmetries of the reduced theory are not
decoupled, because the reduced torsion potential
cannot be simultaneously invariant under both of them.
These anomalous transformation 
properties are at the root of the appearance of the
Chern-Simons terms in the reduced theory, as was
shown by Maharana and Schwarz \cite{mahsch}. 
We recall that the quantity which controls the local dynamics 
of the torsion degrees of 
freedom is the three-form field strength. 
It must turn out to be gauge 
invariant in the reduced theory 
- because, after all,
it will have been derived from 
the original, higher-dimensional theory which
is anomaly-free. In fact, one can  
easily define such a gauge-invariant 
three-form torsion field strength 
in terms of the reduced quantities, starting
with the completely antisymmetric 
derivative of any of the above two torsion
potentials, and then adding the 
gauge pieces as needed to cancel the anomaly. 
Further, in order to treat the anomalous transformation
properties of the torsion after the dimensional descent in a
manifestly symmetric manner, following 
\cite{mahsch} one can define
the reduced torsion by $B_{\mu\nu} = \bar B_{\mu\nu} - (1/2)
(W_{\mu} V_{\nu} - W_{\nu} V_{\mu})$. Then, the gauge-invariant
field strength can be written as $H=dB - (1/2) WdV - (1/2)VdW$,
using the form notation. As we will see below, this approach
enforces manifest $T$-duality invariance at the one-loop
level, and hence is the correct stepping stone towards
extending duality to two loops and beyond.
It is then easy to check that the
explicit gauge dependence of $H$ precisely cancels the
gauge variation of $dB$. As we will see below, the 
reduced torsion actually appears in the dimensionally
reduced action in this form, and hence this field strength 
is indeed the correct choice for the
reduced dynamical variable.
To illustrate this, we will now carry out the 
reduction of (\ref{actnor}) in a step-by-step
fashion. In that way, 
we will verify that the three-form torsion field 
picks up exactly the gauge 
field couplings prescribed here. Let us 
now review this, and 
complete the reduction.

With the ansatz (\ref{dans}), 
one can readily find the tangent space curvature
forms. Here we give a detailed account 
of the results because we will need
them later, when we turn our 
attention to the two-loop corrections.
All the quantities are manifestly 
gauge-invariant when we choose to
work with the basis forms 
$e^a = e^a{}_{\mu} dx^{\mu}, e^y = \exp(\sigma) 
(dy + V_{\mu} dx^{\mu})$, where
the vielbein are defined with 
respect to the reduced metric:
$g_{\mu\nu} = \eta_{ab} e^{a}{}_{\mu} e^b{}_{\nu}$. 
The vielbein matrix $\{e^a{}_{\mu}\}$
defines the transformation law between 
the tangent basis and the holonomic basis
$\{dx^{\mu}\}$. 
We will also need the unreduced vielbein, denoted
by $\bar e^a{}_A, \bar e^y{}_A$, and defined
by the unreduced metric. 
Here the capital latin indices run over
all the coordinates, $\{x\}$ and $y$. 
It is then easy to verify that $\bar e^{a}{}_{\mu} =
e^{a}{}_{\mu}$, $\bar e^{a}{}_{y} =0$,
$\bar e^{y}{}_{\mu}=\exp(\si) V_{\mu}$ 
and $\bar e^{y}{}_{y}=\exp(\si)$.
The tangent space curvature
components then are
\begin{eqnarray} \label{curcomp}
\bar R_{ayby} &=& \frac{1}{4} 
e^{2 \sigma} Z_{ab} -  \Theta_{ab}
\nonumber \\
\bar R_{aybc} &=& -\frac{1}{2} 
e^{\sigma} \Bigl( \nabla_a V_{bc}  + 
2 \nabla_a \sigma V_{bc} + 
2 \nabla_{[c} V_{b]a} \Bigr) \\
\bar R_{abcd} &=& R_{abcd} 
- \frac{1}{4} e^{2 \sigma} \Bigl(
\frac{1}{2} V_{ac} V_{bd} 
- \frac{1}{2} V_{ad} V_{bc}  + V_{ab} V_{cd} \Bigr) 
\nonumber
\end{eqnarray}
The overbar
distinguishes between the 
unreduced and the reduced curvature.
We note that after dimensional reduction we can
specify a tensor ${\cal T}$ by either the tangent
space components $T^{a_1 ... a_p}{}_{b_1 ... b_q}$
or holonomic components 
$T^{\mu_1 ... \mu_p}{}_{\nu_1 ... \nu_q}$,
with the vielbein matrix $e^{a}{}_{\mu}$ defining the
change of basis.
We introduce the symmetric tensors 
\be
Z_{\mu\nu}=V_{\mu\lambda}V_{\nu}{}^{\lambda}\ \ \ \ \ 
T_{\mu\nu}=W_{\mu\lambda}W_{\nu}{}^{\lambda}\ \ \ \ \ 
\Theta_{\mu\nu} = \exp(-\sigma) 
\nabla_{\mu} \nabla_{\nu} \exp(\sigma)
\ee
as a useful shorthand notation. 
The quantities $Z$, $T$ and 
$\Theta$ will be the traces of these tensors.
The antisymmetric tensors $V_{\mu\nu} = 
\partial_{\mu} V_{\nu} - \partial_{\nu} V_{\mu}$ and 
$W_{\mu\nu} = \partial_{\mu} W_{\nu} - 
\partial_{\nu} W_{\mu}$ are the standard 
gauge field strengths of the $U(1)$ 
gauge fields $V_{\mu}$ and $W_{\mu}$, respectively. 
The metric-dilaton sector of the action 
(\ref{actnor}) is now very easy to obtain.
All we need is simply to recall that 
the Ricci scalar is basis-independent, and
trace out the tangent space curvature 
(\ref{curcomp}). The result can be brought in the
standard form after substituting the 
definition of the reduced dilaton in terms
of the original one and the modulus 
$\sigma$, and carrying out a partial integration
of some of the terms \cite{mahsch}. 
The reduction of the three-form
kinetic term is slightly subtler 
because of the anomaly. On the tangent space,
$\bar H^2_{\mu\nu\lambda} = \bar H^{T~2}_{ABC}$. 
Writing out the $y$ indices explicitly,
one finds $\bar H^{T~2}_{ABC} = \bar H^{T}_{abc} \bar H^{T~abc} 
+ 3 \bar H^{T}_{aby} \bar H^{T~aby}$.
But, by the decomposition of the two-form torsion,
$\bar H_{\mu\nu y} = 3 \partial_{[\mu} \bar B_{\nu y]} 
= \partial_{\mu} W_{\nu} - \partial_{\nu} W_{\mu} = W_{\mu\nu}$. 
This yields
\be
\label{tortsy}
\bar H^T_{aby} = \bar e_a{}^{A} \bar e_b{}^{B} 
\bar e_{y}{}^{C} \bar H_{ABC} = 
\exp(-\sigma) e_a{}^{\mu} e_b{}^{\nu} 
\bar H_{\mu\nu y} = \exp(-\sigma) W_{ab}
\ee
As a result, $\bar H^T_{aby} \bar H^{T~aby} 
= \exp(-2 \sigma) W_{ab} W^{ab} = 
\exp(-2 \sigma) W_{\mu\nu} W^{\mu\nu}$. 
Now, the tangent space expression
for $\bar H^T_{abc}$ is
\begin{eqnarray}
\label{tortsst}
\bar H^T_{abc}&=& \bar e_{a}{}^{A} \bar e_{b}{}^{B} 
\bar e_{c}{}^{C} \bar H_{ABC} \nonumber \\
&=& e_{a}{}^{\mu} e_{b}{}^{\nu} 
e_{c}{}^{\lambda} \bar H_{\mu\nu\lambda} + 3 
e_{[a}{}^{\mu} e_{b}{}^{\nu} \bar e_{c]}{}^{y} 
\bar H_{\mu\nu y} \\
&=& e_{a}{}^{\mu} e_{b}{}^{\nu} e_{c}{}^{\lambda}
\Bigl( \bar H_{\mu\nu\lambda} 
- 3 V_{[\lambda} W_{\mu\nu]} \Bigr) \nonumber 
\end{eqnarray}
where $\bar H_{\mu\nu\lambda} = 3 \partial_{[\mu} \bar B_{\nu\lambda]} 
= \nabla_{\mu} \bar B_{\nu\lambda}
+ cyclic~permutations$. The pull-back of the 
reduced tangent space three-form
$\bar H^T_{abc}$ to the holonomic basis 
is the sought-after gauge-invariant
reduced torsion field strength. 
For note that when we define 
\be
\label{giH}
H_{\mu\nu\lambda} = 
\bar H_{\mu\nu\lambda} - 3 W_{[\mu\nu} V_{\lambda]}
\ee
the gauge-dependent pieces in 
it are invariant under 
$W_{\mu} \rightarrow W_{\mu} 
+ \partial_{\mu} \lambda_y$ but change under the
translations along $y$ according to 
$\delta \Bigl(- 3 W_{[\mu\nu} V_{\lambda]}\Bigr) = 
- 3 W_{[\mu\nu} \partial_{\lambda]} \omega = 
- W_{\mu\nu} \partial_{\lambda} \omega 
+ cyclic~ permutations$. This cancels 
the gauge variation of 
$3 \partial_{[\mu} \bar B_{\nu\lambda]}$, as can be 
readily verified. Thus $H_{\mu\nu\lambda}$ is precisely
the gauge-invariant torsion which we have 
discussed at the end of the previous paragraph! 
The gauge anomaly in it 
takes a slightly unusual guise because 
it is not symmetric in $V$ and $W$.
This, as we have explained above, is because the torsion 
components $\bar B_{\mu\nu}$ are
invariant under the reduced Kalb-Ramond $U(1)$ 
gauge symmetry, and not invariant under the Kaluza-Klein
gauge group, generated by the translations along
$y$. But as the Bianchi 
identity for $H_{\mu\nu\lambda}$
yields $\partial_{[\rho}  H_{\mu\nu\lambda]} 
\propto W_{[\rho\mu} V_{\nu\lambda]}$,
we can see that the dynamical role of the 
two gauge fields in the anomaly is
equivalent. An egalitarian description of 
the two symmetries is readily available;
we have already introduced it above.
All one needs to do is to refer to a 
different form of the reduced two-form torsion, 
mentioned before:
\be
\label{fintor}
B_{\mu\nu} = \bar B_{\mu\nu} 
- \frac{1}{2} \Bigl( W_{\mu} V_{\nu} - 
W_{\nu} V_{\mu} \Bigr)
\ee
In terms of this potential, the gauge-invariant 
torsion field strength becomes
\be
\label{fintorH}
H_{\mu\nu\lambda} = \nabla_{\mu} 
B_{\nu\lambda} 
- \frac{1}{2} W_{\mu\nu} V_{\lambda} 
- \frac{1}{2} V_{\mu\nu} W_{\lambda}
+ cyclic~permutations
\ee
which is the component notation for the torsion
three-form of the previous paragraph.
In this form, the torsion is manifestly $T$-duality
invariant, which should occur if duality is to be
an exact symmetry of the full quantum theory of
strings. 
Consequently, $H^{2~T}_{ABC} = 
H_{\mu\nu\lambda} H^{\mu\nu\lambda} 
+ 3 \exp(-2 \sigma)T$. 
This is the last step needed 
in order to find the reduced action; 
it is (after dividing $\Gamma^{(0)}_R$ by
the $y$-volume $\int dy$):
\be
\Gamma^{(0)}_R = \int d^{d}x \sqrt{g}~ e^{-2\phi}
\left\{R+4\dftwo-\dstwo-\frac14\ets Z
-\frac14\emts T-\frac{1}{12} H^2\right\}.
\label{actl}
\ee
with $H$ given in (\ref{fintor}).
The $T$-duality map is (apart from trivial rescalings) 
the transformation 
$\sigma \leftrightarrow -\sigma$, 
$V_{\mu} \leftrightarrow W_{\mu}$, 
and it is obvious that the action
(\ref{actl}) is invariant under it. 
The equations of motion which are obtained 
from varying the action (\ref{actl}) 
(and are simply related to the string $\beta$-functions,
in this order) are covariant under $T$-duality: the 
$\beta$- functions of the dilaton, reduced metric and torsion 
are symmetric, the $\beta$-function of the modulus is
antisymmetric, and the 
$\beta$-functions of the gauge fields
get interchanged, 
as expected from the world-sheet $\sigma$-model
realization of $T$-duality as a map which exchanges
the momentum and winding modes.

An extension of this action 
for the case of $n$ Abelian isometries, and
for general backgrounds which 
admit nontrivial moduli and gauge fields
from both the metric and two-form 
torsion, takes the following form:
\begin{equation}
\label{fact}
S = \int d^{d+1-n} x \sqrt{{g}} e^{- 2\phi} \Bigl\{{R}
 + 4({\nabla} \phi)^2 + \frac{1}{8} Tr \bigl({\cal L}
 \nabla {\cal M}\bigr)^2
- \frac{1}{4} {\cal F}^T_{\mu\nu} 
{\cal LML} {\cal F}^{\mu\nu}
- \frac{1}{12} {H}^2_{\mu\nu\lambda}\Bigr\}
\end{equation}
The capital $T$ denotes matrix transposition.
The $\sigma$-model fields ${\cal M}$ appear after
rearranging the scalar moduli fields.
The correspondence is given by
\begin{equation}\label{w4}
{\cal M}~=~\pmatrix{
G^{-1}&-G^{-1}B \cr
BG^{-1}&G - B^{T}G^{-1}B \cr} ~~~~~~~~~~
{\cal L}~=~\pmatrix{~0~&{\bf 1} \cr
{\bf 1}&~0~\cr}
\end{equation}
where $~G~$, $~B~$ and ${\bf 1}$ 
are $~n \times n~$ matrices
built out of the scalar moduli from the metric and the
axion:$~G~=~\bigl(G_{MN}\bigr)$ and $~B~=~\bigl(B_{MN}\bigr)$.
Here the latin indices $\{M,N\}$ run over the internal space.
The gauge fields are arranged in the
form of the vector multiplet according to
\begin{equation}\label{vectm}
{\cal A}_{\mu}~=~\pmatrix{V^A{}_{\mu} \cr
                          W_{\mu A}\cr}
{}~~~~~~~~~~
{\cal F}_{\mu\nu}~=~\pmatrix{   V^A{}_{\mu\nu} \cr
                                W_{\mu\nu A} \cr}
\end{equation}
and the torsion field strength can be rewritten as
\begin{equation} \label{axfsin}
 {H}_{\mu\nu\lambda} = \nabla_{\mu} 
{B}_{\nu\lambda}
- \frac{1}{2}
{\cal A}^T_{\mu} {\cal L} {\cal F}_{\nu\lambda} +
cyclic ~permutations
\end{equation}
Note
that ${\cal M}^{T} = {\cal M}$ and
${\cal M}^{-1} = {\cal L}{\cal M}{\cal L}$. 
Thus we see
that ${\cal M}$ is a symmetric 
element of $O(n,n,R)$. Therefore an
$O(n,n,R)$ rotation ${\cal M} \rightarrow 
\Omega {\cal M} \Omega^{T}$
and ${\cal F} \rightarrow \Omega {\cal F}$, while
changing ${\cal M}$ and ${\cal F}$, is a symmetry
of the action and the equations of motion.
This symmetry is a generalization of the 
$T$-duality symmetry we have reviewed
above \cite{hassen,mahsch}. One must bear in 
mind, however, that the full
$O(n,n,R)$ group also contains residual 
gauge degrees of freedom - namely,
those diffeomorphisms and gauge transformations of the moduli
inherited from the reduction of the metric and torsion.
These transformations are the elements of
$O(n,n,R)$ which, viewed in terms of 
$d\times d$ blocks, are block-diagonal
and do not mix different tensors but 
merely rearrange their components. It is easy
to see that the net effect is identical 
to the residual global diffeomorphisms 
of the internal space. Thus the nontrivial 
part of the $O(d,d,R)$ group is the
coset $O(d,d,R)/O(d,R) \times O(d,R)$. 
Quite similarly, in the case of a single 
isometry which we will study 
here, the continuous group $O(1,1,R)$ in fact degenerates
down to the $2$-element group 
${\bf Z_2}$ (in this case, we take the group
$O(1,1,R)$ and mod out all the 
rescalings of the compact coordinate - thus
effectively completely determining 
the value of the modulus by the initial 
value problem). Hence, despite its 
appearance, the only nontrivial
$T$-duality symmetry of relevance 
to our problem is the inversion we have
discussed at the end of the previous paragraph.

At this point, we are ready to 
discuss the two-loop corrections. In the next
section, we will present a specific 
form of the $O(\alpha')$ terms and, 
assuming a background with a single 
isometry, carry out dimensional reduction.
The resulting action, as we will see, 
will contain several terms which will
appear to violate $T$-duality, in 
addition to many terms that won't. Our focus
will be on those noninvariant terms,
and we will show that they can be reinterpreted
as the $O(\alpha')$ corrections of the
lowest order duality map. Before we go on,
we would like to note that to the lowest order,
the fields which change under the lowest order
$T$-duality are $\si$ (which changes sign),
and $V_{\mu}$ and $W_{\mu}$ (which get interchanged).
In contrast, the reduced metric $g_{\mu\nu}$,
dilaton $\phi$ and torsion $H_{\mu\nu\lambda}$ 
are invariant - i.e. they are duality-singlets.
While we would naively expect that they should
remain unaffected by duality in higher orders
of the $\alpha'$ expansion as well, we should
observe that the Chern-Simons gauge terms in 
the torsion may in fact induce nontrivial 
transformation corrections to it. We will return
to this issue later.

\section{Reduction of the Two-loop Action}
\setcounter{equation}{0}

In this section, we will present the
two-loop corrections, and perform their
dimensional reduction on the isometry.
A complicating factor in our 
calculations is the presence of the
field redefinition ambiguity. Namely, it
is well-known that the exact 
form of the $O(\alpha')$ corrections is not
unique, but rather that it depends on 
the subtraction scheme adopted in
string field theory. This ambiguity arises
in the following way. The effective action to 
any given order in the $\alpha'$ expansion 
can be obtained in two ways. One of them is to
reverse-engineer the
$\beta$-functions of the world-sheet $\si$-model to
get the action whose extrema 
correspond to the fixed points of the $\si$-model.
However, the
$\beta$-functions are not unique 
beyond the lowest order, since the 
renormalization scheme influences the form of the
$\alpha'$ corrections. 
The renormalization scheme dependence
of the $\beta$-functions is seen as some field redefinition of the
effective action, meaning that configurations
related by this field redefinition are in fact physically 
identical. 
The other way to obtain the effective action is to 
calculate relevant string amplitudes and expand them to the
required order in $\alpha'$. In this approach any 
local field redefinition that
respects all the symmetries of the theory and 
is analytic in $\alpha'$ (and is equal to identity
in the limit $\alpha'=0$), yields
the same on-shell amplitude, so the ambiguity
in the effective action is in this case even greater than in the 
$\beta$-function approach.
``On-shell'', i.e. 
when both the $\beta$-functions and the functional
derivatives of the action vanish, the two descriptions are believed to
coincide. 
The equivalence conjecture simply 
states that there exist a renormalization scheme 
and a field redefinition 
such that the two sets of solutions 
(constructed from the requirement of vanishing of the
$\beta$-functions and the equations of motion) are 
identical.

Therefore, the appearance of 
the $O(\alpha')$ corrections in the
effective action can
be changed by finite renormalizations 
of the string world-sheet 
couplings \cite{tm,gs}. 
In a sense, this blurs the 
notion of string theory quantities as functions
of $\alpha'$ - instead of a 
single set of solutions, one ends up with equivalence
classes, specified
by the field redefinitions. 
Nevertheless, in order to do any 
calculations one has to adopt a concrete
scheme, thus fixing the form 
of the counterterms in the effective action
and the functional dependence 
on $\alpha'$. 
While many of the counterterms 
are affected by field redefinitions, 
a number of them are fixed by the string 
scattering amplitudes. One could 
in principle imagine an analogue of the 
field-theoretic minimal subtraction scheme
by simply retaining only the 
ambiguity-free terms in the action. The merit
of such an approach is that it may lead to 
calculational simplifications for ``on-shell"
configurations.
However, ``off-shell'', i.e. for arbitrary field
configurations, the connection between the $\beta$-functions
and the functional derivatives of the effective action
is a subtle issue. A relation similar to
that ``on-shell" has been suggested in \cite{zam},
but its validity is far less obvious
(although it holds in the known examples - see \cite{tm}).
In \cite{mavmir,jj} relations of this type were used to get
the unique ``off-shell" action starting from the 
known two-loop $\beta$-functions. Specifically,
because the $\beta$-functions are generally more
constrained, and hence less dependent on the redefinitions 
of fields than the functional derivatives of the
effective action, one can use the redefinitions
to change the form of the functional derivatives
until they are a local linear combination of 
the $\beta$-functions. This exausts the field 
redefinition ambiguities, and specifies the two-loop
action uniquely - albeit a non-minimal one. 
Choosing such a scheme hence
is perfectly legitimate, although it may appear 
less appealing initially. But we should underline
that while the equations of motion in this scheme
may appear far more cumbersome than in the above
described ``minimal" scheme, the two-loop effective
action is defined so that the $\sigma$-model
$\beta$-functions are linear combinations of  the functional
derivatives of the effective action - without constraining the
latter to vanish. Thus starting from this action, 
one can easily write down the complete set of 
$\beta$-functions - regardless of whether they
are ``on-shell" or ``off-shell" (i.e. at or away 
from the conformal point).

Alternatively, a scheme may be specified by simultaneously
requiring manifest unitarity in perturbation theory and linear
realization of duality \cite{km}. 
These two requirements have been shown
to also specify the action uniquely, at least on the
special class of cosmological backgrounds that they 
were applied to. Surprisingly, the action obtained in this
way turned out to be identical to the nonminimal
action we have described above - that is, it was 
precisely the action encoding the local relationship
between the $\beta$-functions and the functional derivatives!
It was further observed that 
when one wants to exhibit $T$-duality 
of the action one makes field
redefinitions until the functional 
derivatives coincide with the 
$\beta$-functions. This relationship between the
$\beta$-functions and $T$-duality 
was noted in \cite{km} 
as a very remarkable fact. 

The investigation of \cite{km} was done for
backgrounds with all but one cyclic coordinates,
which were shown to possess the 
full $O(d,d)$ symmetry to two loops. The
effects of the momentum and 
winding modes did not arise there because of the 
symmetry. One can quickly verify 
that both gauge fields are identically zero
on such backgrounds (modulo diffeomorphisms 
and torsion gauge transformations, and provided
that none of the isometry generators are null). 
Our approach here will be based 
on the same scheme as described in
\cite{km}. Here we will extend the analysis
to when the gauge degrees of freedom
are nontrivial. The effective action to two loops is 
(we use the same form of the action as the one derived in
\cite{km} since it also exhibits manifest $T$-duality
in the present case)
\ba
\label{twola}
\Gamma&=&\int d^{d+1}x \sqrt{\bar g}e^{-2\bar \phi}
\left\{\bar R(\bar g)+4(\bar \nabla \bar \phi)^2 
-\frac1{12}\bar H^2\right.
\nonumber\\
&&+\alpha'\lz\left[
-\bar R^2_{GB}+16\left(\bar R^{\mu\nu}-
\frac12\bar g^{\mu\nu}\bar R\right)\bar \nabla_{\mu}\bar \phi
\bar \nabla_{\nu}\bar \phi 
-16\bar \nabla^2\bar \phi (\bar \nabla \bar \phi)^2
+16(\bar \nabla \bar \phi)^4\right.\nonumber\\
&&+\frac12\left(\bar R_{\mu\nu\lambda\rho} 
\bar H^{\mu\nu\alpha} \bar H^{\lambda\rho}{}_{\alpha}
-2 \bar R^{\mu\nu}\bar H^2_{\mu\nu}
+\frac13 \bar R \bar H^2\right)
-2\left(\bar \nabla^{\mu}\bar \nabla^{\nu}
\bar \phi \bar H^2_{\mu\nu}
-\frac13 \bar \nabla^2 \bar \phi
\bar H^2\right)\nonumber\\
&&\left.\left. -\frac23 \bar H^2 
(\bar \nabla \bar \phi)^2
-\frac1{24} \bar H_{\mu\nu\lambda} \bar H^{\nu}{}_{\rho\alpha} 
\bar H^{\rho\sigma\lambda} \bar H_{\sigma}{}^{\mu\alpha}  +
\frac18 \bar H^2_{\mu\nu} \bar H^2{}^{\mu\nu}
-\frac1{144}(\bar H^2)^2\right]\right\}.
\ea
where we introduce more useful shorthand notation:
\be
\bar H^2_{\mu\nu}=\bar H_{\mu\alpha\beta}
\bar H_{\nu}{}^{\alpha\beta},
\ \ \ \ {\rm and} 
\ \ \ \ \ \bar H^2=\bar H_{\mu\alpha\beta}
\bar H^{\mu\alpha\beta}
\ee
The parameter $\lambda_0$ allows us to move between different
string theories: $\lambda_0 = -1/8 $ 
for heterotic, $-1/4$ for bosonic, and $0$ for 
superstring. In this sense, our 
calculations are completely general (although, 
they are of course trivial in the case of the superstring 
- where the $O(\alpha')$ terms vanish identically 
and hence the lowest-order $T$-duality does not acquire any corrections;
curiously, in this context $T$-duality is not a map between the 
solutions of a theory but instead a map between different theories).
The $\bar R-\bar H$ terms include also the 
Lorentz-Chern-Simons terms which
emerge in the heterotic theory to two loops.
The curvature squared terms are 
arranged in the Gauss-Bonnet combination,
$\bar R^2_{GB} = \bar R^2_{\mu\nu\lambda\sigma} 
- 4 \bar R^2_{\mu\nu} + \bar R^2$,
and hence are manifestly unitary\footnote{Manifest
unitarity here may be viewed more as another
curiosity 
than a necessity. It is not necessary to 
require that the effective action (\ref{twola}) be 
manifestly unitary because it represents only an
approximation to the complete theory of strings.
One could have instead resorted to a different scheme
where the curvature corrections, and other terms,
do not appear in the unitary form, keeping in mind
that in the region of the phase space where unitarity is 
violated the approximations made to truncate the full theory also
break down. This merely indicates that in such limits higher 
order corrections must be taken
into account. But as we have
explained in the text, the
action (\ref{twola}) also has the property that
its functional derivatives are locally related to the
$\beta$-functions. It is intriguing to think that
unitarity and $T$-duality are equivalent to 
this reproduction of ``off-shell" $\beta$-functions
at some deep level.}.

While dimensional reduction of this action may appear 
unsavoury at first, it is manageable when carried out 
on the tangent space. As in the one-loop case, 
enforcing reduced gauge symmetries eliminates 
many of the complicated terms 
which appear at intermediate steps 
but cancel in the final answer.
Also, we recall that because the action is a 
scalar, it doesn't matter in which basis
it is written. Therefore, all we 
need to do is to treat the components of
tensors in (\ref{twola}) as tangent 
space quantities and then decompose
them into space-time and internal 
space subsets. While the calculations
are simple, the resulting reduced 
tensors are defined precisely in terms
of the reduced tangent space 
quantities. The only note of warning is that one 
has to carefully work out the 
terms explicitly containing covariant derivatives,
reduce them by splitting the 
connexion, and then pull them back on the
tangent space. This in fact is an application
of the Young tableaux method for 
the reduction of irreducible representations 
of groups - we are reducing 
$Gl(d,1,R)$-invariant quantities to 
$Gl(d-1,1,R)$-invariant ones (or in 
fact $Gl(d,R)$-invariant terms - for
if one views $T$-duality as a solution 
generating technique, one may assume
$p$-brane type backgrounds on which 
the cyclic coordinate is timelike). 
After all this has been accomplished, 
and the cyclic coordinate $y$ integrated 
out, yielding the reduced action, 
we can simply replace 
the reduced tangent space indices with
the holonomic ones.

Thus, the
Gauss-Bonnet sector of the action
yields, after dividing by $l_y = \int dy$ 
and discarding the boundary terms,
\ba
\label{gaussb}
(1/l_y)\bar R^2_{GB} &=& e^{\sigma}
\Bigl\{ R^2_{GB} +\frac{3}{16} e^{4\sigma}Z^2 
- \frac{3}{8} e^{4\sigma}Z^2_{\mu\nu} 
- 4 R \Theta  + 8 R_{\mu\nu} \Theta^{\mu\nu} \nonumber \\
&&+ 4 e^{2\sigma}R_{\mu\nu} Z^{\mu\nu} 
- \frac{1}{2} e^{2\sigma} R Z - 
e^{2\sigma} R_{\mu\nu\lambda\sigma} 
\Bigl(V^{\mu\nu} V^{\lambda\sigma} 
+ V^{\mu\lambda} V^{\nu\sigma} \Bigr) \nonumber \\
&&+ 3 e^{2 \sigma} \Theta Z - 
6 e^{2 \sigma} \Theta_{\mu\nu} Z^{\mu\nu} 
+ e^{-2\sigma} \nabla_{\mu}(e^{2\sigma} 
V_{\nu\lambda})\nabla^{\mu}(e^{2\sigma} V^{\nu\lambda}) \\
&&-2e^{-2\sigma} \nabla_{\mu}(e^{2\sigma} 
V^{\mu\nu})\nabla^{\lambda}(e^{2\sigma} V_{\lambda\nu}) + 
4 V^{\mu\nu} \nabla_{\nu}(e^{2\sigma} V_{\mu\lambda}) 
\nabla^{\lambda} \sigma \nonumber \\
&&-4 V^{\mu\nu} \nabla^{\lambda}(e^{2\sigma} V_{\mu\lambda}) 
\nabla_{\nu} \sigma + 
2 e^{2\sigma} (\nabla \sigma)^2 Z 
- 4 e^{2\sigma} \nabla_{\mu} \sigma \nabla_{\nu} \sigma
Z^{\mu\nu} \Bigr\} \nonumber
\ea
Next, we notice that the dilaton
derivatives are handled with ease. Following 
the route described in the previous 
paragraph, we find that the reduction formulae 
for the derivatives of the dilaton, 
pulled back to the tangent space, read
\be
\label{dderiv}
\bar \nabla_y \bar \nabla_y \bar \phi = \nabla_{a} \sigma 
\nabla^{a} \bar \phi \ \ \ \ \
\bar \nabla_a \bar \nabla_y \bar \phi = -\frac{1}{2} 
e^{\sigma} V_{ab} \nabla^{b} \bar \phi \ \ \ \ \
\bar \nabla_a \bar \nabla_b \bar \phi = \nabla_a \nabla_b \bar \phi 
\ee 
Combining these and the expressions 
(\ref{curcomp}) for the tangent space
curvature components, we immediately 
get the expression for the reduced
dilaton-curvature couplings $\bar R-\bar \phi$
(which are given by 
$16(\bar R_{\mu\nu} - (1/2) \bar g_{\mu\nu} \bar R) 
\bar \nabla^{\mu} \bar \phi
\bar \nabla^{\nu} \bar \phi$):
\ba
\label{dcurv}
(1/l_y) (\bar R-\bar \phi) &=& 16 \Bigl(R_{\mu\nu} 
- \frac{1}{2} g_{\mu\nu} R \Bigr) \nabla^{\mu} \Phi 
\nabla^{\nu} \Phi 
- 16 \Bigl(\Theta_{\mu\nu} 
- g_{\mu\nu} \Theta  \Bigr) \nabla^{\mu} \Phi 
\nabla^{\nu} \Phi \nonumber \\
&& - 8 \Bigl(Z_{\mu\nu} 
- \frac{1}{4} g_{\mu\nu} Z \Bigr) \nabla^{\mu} \Phi 
\nabla^{\nu} \Phi 
\ea 
Similarly, the curvature-torsion couplings ($\bar R-\bar H$) 
in the original action (given by the expression $\bar R-\bar H = (1/2)
(\bar R_{\mu\nu\lambda\sigma} 
\bar H^{\mu\nu\rho} \bar H_{\lambda\sigma}{}_{\rho} 
- 2 \bar R_{\mu\nu} \bar H^{2~\mu\nu}
+(1/3) \bar R \bar H^2$)) reduce to
\ba
\label{RHred}
(1/l_y) (\bar R-\bar H) &=& \frac{1}{2}\Bigl\{R_{\mu\nu\lambda\sigma} 
 H^{\mu\nu\rho} H^{\lambda\sigma}{}_{\rho} 
- 2 R_{\mu\nu} H^{2~\mu\nu}
+\frac{1}{3} R H^2 + 2 \Theta_{\mu\nu} H^{2~\mu\nu} \nonumber \\
&&- \frac{2}{3} \Theta H^2 + e^{-2\sigma} 
R_{\mu\nu\lambda\sigma}  
W^{\mu\nu} W^{\lambda\sigma} - 4 e^{-2\sigma} 
R_{\mu\nu} T^{\mu\nu} + e^{-2\sigma} R T
\nonumber \\
&&-\frac{1}{2} e^{2\sigma} 
H_{\mu\nu\rho} H_{\lambda\sigma}{}^{\rho}
\Bigl(V^{\mu\nu} V^{\lambda\sigma} 
+ V^{\mu\lambda} V^{\nu\sigma} \Bigr) 
+ e^{2\sigma} H^2{}_{\mu\nu} Z^{\mu\nu} \nonumber \\
&&- \frac{1}{12} e^{2\sigma} H^2 Z
- \frac{3}{4} Z T + 3 Z^{\mu\nu} T_{\mu\nu} 
- \frac{1}{2} V^{\mu\nu} V^{\lambda\sigma} 
W_{\mu\lambda} W_{\nu\sigma} \\
&& - \frac{1}{2} V^{\mu\nu} W_{\mu\nu} 
V^{\lambda\sigma} W_{\lambda\sigma} +
2 \nabla^{\mu} V^{\nu\lambda} W_{\mu}{}^{\sigma} 
H_{\nu\lambda\sigma} \nonumber \\
&&-2 \nabla_{\lambda} V^{\mu\lambda} W^{\nu\sigma} 
H_{\mu\nu\sigma}
+ 4 \nabla^{\mu} \sigma V^{\nu\lambda} W_{\mu}{}^{\sigma} 
H_{\nu\lambda\sigma} \nonumber\\
&&+ 4 \nabla^{\lambda} \sigma V^{\nu\mu} W_{\mu}{}^{\sigma} 
H_{\nu\lambda\sigma}
- 6 \nabla_{\lambda}\sigma V^{\mu\lambda} W^{\nu\sigma} 
H_{\mu\nu\sigma}
\Bigr\} \nonumber
\ea
The dilaton-torsion terms $\bar \phi-\bar H$ 
($=(2/3) \bar \nabla^2 \bar \phi \bar H^2 -
2\bar \nabla_{\mu} \bar \nabla_{\nu} 
\bar \phi \bar H^{2~\mu\nu}$) become
\ba
\label{diltorred}
(1/l_y) (\bar \phi-\bar H) &=& \frac{2}{3} \nabla^2 \bar \phi H^2 -
2 \nabla_{\mu} \nabla_{\nu} \bar \phi H^{2~\mu\nu} 
+ \frac{2}{3} \nabla_{\mu} \sigma 
\nabla^{\mu} \bar \phi H^2 \nonumber \\
&&- 2 \nabla_{\mu} \bar \phi V^{\mu\nu} 
W^{\lambda\sigma} H_{\nu\lambda\sigma} 
+ 2 e^{-2\sigma} \nabla^2 T -
4 e^{-2\sigma} 
\nabla_{\mu} \nabla_{\nu} \bar \phi T^{\mu\nu}
\ea
We recall that to reduce $(\bar H^2)^2$ 
we merely need to square the expression for 
$\bar H^2$ which we have obtained in the
previous section, in the course 
of deriving the action (\ref{actl}). 
The only remaining terms whose 
reduction is not trivial are the terms quartic in
the torsion $\bar H$. Our tangent 
space formulas for the torsion allow us
to write down the following two 
expressions for these terms:
\ba
\label{tor41}
(1/l_y) \bar H_{\mu\nu\lambda}
\bar H^{\nu}{}_{\sigma\rho}\bar H^{\sigma\gamma\lambda}
\bar H_{\gamma}{}^{\mu\rho} &=&
 H_{\mu\nu\lambda} H^{\nu}{}_{\sigma\rho}
 H^{\sigma\gamma\lambda}
 H_{\gamma}{}^{\mu\rho} 
\nonumber \\
&&+ 6 e^{-2\sigma}  H_{\mu\nu\rho} 
 H_{\lambda\sigma}{}^{\rho} W^{\mu\lambda}
W^{\nu\sigma} + 3 e^{-4\sigma} T_{\mu\nu} T^{\mu\nu}
\ea
and
\ba
\label{tor42}
(1/l_y) \bar H^2{}_{\mu\nu}\bar H^{2~\mu\nu} &=&
 H^2{}_{\mu\nu} H^{2~\mu\nu} 
+ 2 e^{-2\sigma}  H_{\mu\nu\rho} 
 H_{\lambda\sigma}{}^{\rho} W^{\mu\nu}
W^{\lambda\sigma} \nonumber \\
&&+ 4 e^{-2\sigma}  H^2{}_{\mu\nu} T^{\mu\nu} 
+ 4 e^{-4\sigma} T_{\mu\nu} T^{\mu\nu} + e^{-4\sigma} T^2
\ea 

With this at hand, we can finally 
write down the explicit form of the reduced action
(\ref{actl}). Here we still need to reduce 
the dilaton, using the formula (\ref{dans}). 
Further, we also need to simplify the terms containing
derivatives of the gauge fields. This is accomplished with a
number of partial integrations.
When we rewrite the action in 
terms of the shifted dilaton, after 
some simple algebra, we find the following 
contributions, which for convenience 
we present here as sum of several terms:
\be
\label{finact}
\Gamma^{(2)} = \Gamma^{(2)}_1 + \Gamma^{(2)}_2 
+ \Gamma^{(2)}_3 + \Gamma^{(2)}_4 + \Gamma^{(2)}_5
\ee
where 
\ba
\label{fin1}
\Gamma^{(2)}_1&=&\int d^d x \sqrt{ g}e^{-2 \phi}
\left\{ R+4( \nabla  \phi)^2 
-\frac1{12} H^2\right.
\nonumber\\
&&+\alpha'\lz\left[
- R^2_{GB}+16\left( R^{\mu\nu}-
\frac12 g^{\mu\nu} R\right) \nabla_{\mu} \phi
 \nabla_{\nu} \phi 
-16 \nabla^2 \phi ( \nabla  \phi)^2
+16( \nabla  \phi)^4\right.\nonumber\\
&&+\frac12\left( R_{\mu\nu\lambda\rho} 
 H^{\mu\nu\alpha}  H^{\lambda\rho}{}_{\alpha}
-2  R^{\mu\nu} H^2_{\mu\nu}
+\frac13  R  H^2\right)
-2\left( \nabla^{\mu} \nabla^{\nu}
 \phi  H^2_{\mu\nu}
-\frac13  \nabla^2  \phi
 H^2\right)\nonumber\\
&&\left.\left. -\frac23  H^2 
( \nabla  \phi)^2
-\frac1{24}  H_{\mu\nu\lambda}  H^{\nu}{}_{\rho\alpha} 
 H^{\rho\sigma\lambda}  H_{\sigma}{}^{\mu\alpha}  +
\frac18  H^2_{\mu\nu}  H^2{}^{\mu\nu}
-\frac1{144}( H^2)^2\right]\right\}.
\ea
which we recognize to have the same form as the
original two-loop action (\ref{twola}), except that it
depends on the reduced fields. The remaining terms describe
the gauge- and moduli-dependent corrections. 
The gauge-dependent terms are as follows:
\ba
\label{fin2}
\Gamma^{(2)}_{2}&=&\aplz\int d^d x\sqrt{g}e^{-2\phi}\Bigl\{
-\frac3{16}Z^2 e^{4\si}+\frac1{16}T^2 e^{-4\si}
+\frac38 Z^{\alpha\beta}Z_{\alpha\beta}e^{4\si}
+\frac38 T^{\alpha\beta}T_{\alpha\beta}e^{-4\si}
\nonumber\\
&&
+\frac32 Z^{\alpha\beta}T_{\alpha\beta}-\frac38 ZT 
-\frac14\left(V_{\alpha\gamma}
V_{\beta\mu}W^{\alpha\beta}W^{\gamma\mu}
+V_{\alpha\beta}W^{\alpha\beta}V_{\gamma\mu}W^{\gamma\mu}
\right) \Bigr\}
\ea
gives the sector quartic in gauge fields, while 
the gauge-curvature and gauge-torsion couplings are given by: 
\newpage
\ba
\label{fin3}
\Gamma^{(2)}_{3}&=&\aplz\int d^d x\sqrt{g}e^{-2\phi}\Bigl\{
\frac12 RZ\ets+\frac12 RT\emts
+\frac12R^{\alpha\beta\mu\nu}V_{\alpha\beta}V_{\mu\nu}\ets
\nonumber\\
&&
+\frac12R^{\alpha\beta\mu\nu}W_{\alpha\beta}W_{\mu\nu}\ets
-2R^{\alpha\beta}Z_{\alpha\beta}\ets
-2R^{\alpha\beta}T_{\alpha\beta}\emts 
+\nabla_{\gamma}V^{\alpha\beta}H_{\alpha\beta\rho}W^{\gamma\rho}
\nonumber\\
&&
+\left(\nabla^{\beta}V_{\beta\alpha}\right)
H^{\alpha\rho\gamma}W^{\rho}{}_{\gamma}
-\frac14\ets\left(V_{\alpha\gamma}V_{\beta\mu}
+V_{\alpha\beta}V_{\gamma\mu}\right)
H^{\alpha\beta\rho}H^{\gamma\mu}{}_{\rho}
\nonumber\\
&&
-\frac14\emts\left(W_{\alpha\gamma}W_{\beta\mu}
-W_{\alpha\beta}W_{\gamma\mu}
\right)H^{\alpha\beta\rho}
H^{\gamma\mu}{}_{\rho}
+\frac12 \ets Z^{\mu\nu}H^2_{\mu\nu}
\nonumber\\
&&
+\frac12 \emts T^{\mu\nu}H^2_{\mu\nu}
-\frac1{24}\ets Z H^2-\frac1{24}\emts T H^2 \Bigr\}
\ea
Note that in these expressions there are terms which 
are not symmetric under the
permutation of the gauge fields.

The moduli-dependent selfinteractions and the interactions with the
dilaton field are
\ba
\label{fin4}
\Gamma^{(2)}_{4}&=&\aplz\int 
d^dx\sqrt{g}e^{-2\phi}\Bigl\{
+16[\nabla \phi \nabla \si + \frac14(\nabla \si)^2]
[2 (\nabla \phi)^2 + \nabla \phi \nabla \si + \frac14(\nabla \si)^2]
\nonumber \\
&&16\Theta\left(\pa \phi
+\frac12\pa \si\right)^2 
-16\Theta^{\mu\nu}\left(\pa_{\mu}\phi
+\frac12\pa_{\mu}\si\right)
\left(\pa_{\nu}\phi+\frac12\pa_{\nu}\si\right)
\nonumber\\
&&-16\left(\pa_{\mu}\si\pa^{\mu}\phi+
\frac12(\pa\si)^2+\frac12\nabla^2\si\right)
\left(\pa\phi+\frac12\pa\si\right)^2 \nonumber \\
&&-16\nabla^2\phi [\nabla \phi \nabla \si + \frac14(\nabla \si)^2]
\Bigr\}
\ea
and the interactions of scalars 
with the curvature, torsion and gauge fields
are
\ba
\label{fin5}
\Gamma^{(2)}_{5}&=&\aplz\int 
d^d x\sqrt{g}e^{-2\phi}\Bigl\{-4\left(R^{\mu\nu}
-\frac12g^{\mu\nu}R\right)\pa_{\mu}\si
\pa_{\nu}\si +H^2_{\mu\nu}\pa^{\mu}\si\pa^{\nu}\si
\nonumber\\ 
&&-\frac16(\pa\si)^2H^2
+2\pa^{\alpha}\phi V_{\rho\alpha}W_{\beta\gamma}
H^{\rho\beta\gamma}
+2\left(\pa_{\gamma}\si V_{\alpha\beta}+\pa_{\beta}\si
V_{\alpha\gamma}\right) H^{\alpha\beta\rho}W^{\gamma}{}_{\rho}
\nonumber\\
&&
+2\pa^{\beta}\si V_{\beta\alpha}W_{\rho\gamma}
H^{\alpha\rho\gamma}
+2Z\left(\nabla^2\phi+\frac12\nabla^2\si\right)\ets + 
2T\left(\nabla^2\phi+\frac12\nabla^2\si\right)\emts\nonumber\\
&&-2Z\ets\left(\pa\phi+\frac12\pa\si\right)^2
-2T\emts\left(\pa\phi+\frac12\pa\si\right)^2\nonumber\\
&&-4Z^{\alpha\beta}\ets \nabla_{\alpha}\left(\pa_{\beta} \phi
+\frac12\pa_{\beta}\si\right)
-4T^{\alpha\beta}\emts \nabla_{\alpha}\left(\pa_{\beta} \phi
+\frac12\pa_{\beta}\si\right)
\Bigr\}
\ea

Note that as we have said above, 
the contributions to the 
reduced action which we have listed here
contain both terms which are invariant 
under the one-loop level duality transformations
$\sigma \leftrightarrow -\sigma$, 
$V_\mu \leftrightarrow W_\mu$ and terms which 
are not symmetric under this map. It 
is these latter terms which we are interested in
here. They are the ones forcing us to correct 
the one-loop duality map with $O(\alpha')$
contributions. Fortunately, these terms are 
far fewer than the symmetric ones - hence
making our task viable. For completeness sake
we will present
both classes of terms, before we
proceed to discuss the corrections in the 
next section. In order to separate the
two-loop contributions to (\ref{finact}) into 
one-loop duality-invariant
and duality-noninvariant parts, we can 
apply the transformations,
and work out $\Gamma_{inv} = 
(1/2)(\Gamma_2 + T \Gamma_2)$ and
$\Gamma_{ninv} = (1/2)(\Gamma_2 - T\Gamma_2)$. 
Using this, 
the duality-violating sector of the 
reduced $O(\alpha')$ action is 
\ba
\label{noninvtact}
\Gamma^{(2)}_{ninv}&=&\aplz\int d^d x\sqrt{g}e^{-2\phi}\Bigl\{
-4 \nabla_{\mu} \sigma \nabla^{\mu} (\nabla \sigma)^2 
-\nabla_\mu \sigma \nabla^\mu  
\left[e^{-2\sigma} T +e^{2\sigma} Z\right] 
\nonumber \\
&&+ \frac{1}{8} \left[e^{-4\sigma} T^2 - e^{4\sigma} Z^2\right] 
+\frac14 H_{\alpha\beta\sigma} H_{\mu\nu}{}^{\sigma} 
\left[W^{\alpha\beta}W^{\mu\nu}\emts
-V^{\alpha\beta}V^{\mu\nu}\ets \right]  \\
&&+\left[\nabla^\mu V^{\nu\lambda} W_{\mu}{}^{\sigma} -
\nabla^\mu W^{\nu\lambda} V_{\mu}{}^{\sigma} \right] 
H_{\nu\lambda\sigma}
+\frac12 \left[V^{\nu\lambda} W^{\mu\sigma} -
W^{\nu\lambda} V^{\mu\sigma} \right] 
\nabla_\lambda H_{\nu\mu\sigma} \nonumber \\
&&+2 \nabla_\gamma \sigma V_{\mu\nu} W^{\gamma}{}_{\lambda} 
H^{\mu\nu\lambda}
+2 \nabla_\gamma \sigma W_{\mu\nu} V^{\gamma}{}_{\lambda} 
H^{\mu\nu\lambda} \Bigr\} \nonumber
\ea
and finally, the symmetric part of the action is
\ba
\label{invact}
\Gamma^{(2)}_{inv}&=&\aplz\int 
d^d x\sqrt{g}e^{-2\phi}\Bigl\{
-R^2_{GB}+ \frac{1}{2} \left[\rhh-2R^{\mu\nu}H^2_{\mu\nu} 
+\frac13 R H^2 \right] \nonumber \\
&&+16\left(R^{\mu\nu}-\frac12g^{\mu\nu}R\right)\pa_{\mu}\phi
\pa_{\nu}\phi -\frac1{24}\hfour
+ \frac18H^2_{\mu\nu}H^2{}^{\mu\nu} \nonumber \\
&&-\frac1{144}(H^2)^2 
-2\nabla^{\mu}\nabla^{\nu}\phi H^2_{\mu\nu}
+\frac23H^2\nabla^2\phi 
-\frac23 H^2(\pa\phi)^2 -(\nabla \sigma)^4 
\nonumber \\
&&-16 (\nabla \sigma \nabla \phi)^2 
+ 8 (\nabla \sigma \nabla \phi) 
\nabla^2 \sigma 
+ 4 (\nabla \sigma)^2 \nabla^2 \phi 
-4\left(R^{\mu\nu}-\frac12g^{\mu\nu}R\right)\pa_{\mu}\sigma
\pa_{\nu}\sigma \nonumber \\
&&+\frac38 
\left[Z^{\alpha\beta}Z_{\alpha\beta}e^{4\si}
+ T^{\alpha\beta}T_{\alpha\beta}e^{-4\si} \right]
-\frac{1}{4} V^{\mu\nu} 
V^{\lambda\sigma} W_{\mu\lambda} W_{\nu\sigma}
+\frac12 Z^{\alpha\beta}T_{\alpha\beta} \\
&&-\frac18 ZT
+\frac12 \left[RZ\ets+RT\emts \right]
-2R^{\alpha\beta} \left[Z_{\alpha\beta}\ets
+T_{\alpha\beta}\emts
\right] \nonumber \\
&&+\frac12R^{\alpha\beta\mu\nu} 
\left[V_{\alpha\beta}V_{\mu\nu}\ets
+W_{\alpha\beta}W_{\mu\nu}\emts \right] - 
4 \nabla_\mu \nabla_\nu \phi 
\left[Z^{\mu\nu}\ets+T^{\mu\nu}\emts\right]\nonumber \\
&&+\left[2 \nabla^2 \phi 
- 2(\nabla \phi)^2 -\frac12 (\nabla \sigma)^2 \right]
\left[Z\ets+T\emts\right] 
- \frac1{24} H^2 \left[Z\ets+T\emts\right]
\nonumber \\
&&+\frac12 H^2_{\mu\nu} 
\left[Z^{\mu\nu}\ets+T^{\mu\nu}\emts\right]
-\frac14 H_{\alpha\beta\sigma} H_{\mu\nu}{}^{\sigma} 
\left[V^{\alpha\beta}V^{\mu\nu}\ets
+W^{\alpha\beta}W^{\mu\nu}\emts \right] \nonumber \\
&&+2 \nabla_\gamma \sigma V_{\mu\nu} W^{\nu}{}_{\lambda} 
H^{\mu\gamma\lambda} \Bigr\} \nonumber
\ea
We will not need (\ref{invact}) for the remainder of
our calculations. We note however that it is an integral 
part of the subtraction scheme which we have adopted when
taking the action (\ref{twola}), and it does correspondingly
affect the appearance of the solutions to $O(\alpha')$.

\section{Corrections to $T$-duality}
\setcounter{equation}{0}

We can finally embark on the 
study of the corrections 
(\ref{noninvtact}) and their 
reinterpretation as the 
modification of the one-loop level 
$T$-duality map. In order to do
so, we will approach the problem 
from a perturbative standpoint. Namely, 
the full $O(\alpha')$ action we 
have started with is, generically,
a good approximation to string 
dynamics only to order $O(\alpha')$.
This seemingly obvious statement 
in fact renders valuable information
for the understanding of the role 
of (\ref{noninvtact}). Given the
noninvariant terms (\ref{noninvtact}) 
and the one-loop level duality
$\sigma \leftrightarrow - \sigma, 
V_{\mu} \leftrightarrow W_{\mu}$,
the natural way to reconcile 
them is to interpret
(\ref{noninvtact}) as the $O(\alpha')$ 
terms in the expansion of the
exact $T$-duality map, which presumably 
exists in a complete 
quantum theory which admits various 
string theories as its limits.
Hence, one ought to be able 
to incorporate these terms by 
redefining the duality-invariant 
one-loop action - in effect shifting
the one-loop level fields by 
amounts proportional to $\alpha'$,
while preserving any 
other symmetries the one-loop level
theory has. With this in mind, 
we only need to ensure that the
noninvariant terms (\ref{noninvtact}) 
be absorbed to $O(\alpha')$.
Any further deviation away from 
the perturbative form of 
$T$-duality, at any higher loop order,
that these terms might participate 
in, can be safely ignored from 
the point of view of the effective 
action, for we must take into account the 
influence of other terms beyond 
the truncation we adopt here, 
to consistently address such 
higher order effects. While our
approach may appear limited to 
the two loop level, it conforms
with the idea that $T$-duality is 
perturbatively exact. As we will show
below, the duality-violating 
corrections (\ref{noninvtact}) will be
consistently reincorporated 
in the one-loop action, respecting all other
symmetries, and in particular 
preserving the form of the one-loop
gauge anomaly contributions. 
At the next order, $\alpha'^2$, we
expect that the terms 
violating the two-loop form of duality, that
will emerge from our calculations,
can be similarly absorbed away
as further corrections of the 
$O(\alpha'^0)$ and $O(\alpha')$
sectors of the action. The change these terms
will induce in the duality map 
will be quadratic in $\alpha'$. Hence
such terms can arise as the 
subleading corrections to the two-loop
duality-preserving contributions 
in (\ref{invact}) and 
as $O(\alpha'^2)$ corrections of 
the one-loop action (\ref{actl}). Given 
the proliferation of the 
duality-preserving terms, we expect that
such analytical cancellations 
will go on {\it ad infinitum}, yielding a
Taylor-series expression for 
the complete $T$-duality map.  Note that
this does not imply that all 
the physical variables of string
theory must also admit such 
Taylor-series expansions in $\alpha'$.
While this may certainly be 
the case for some of the string degrees
of freedom, all we can deduce 
from the proposed realization of the exact
$T$-duality is that the map will 
not introduce any new $\alpha'$ 
poles in the image of the 
background it acts on. This means that $T$-duality 
relates backgrounds it acts on 
such that both the original 
and the image admit similar $\alpha'$
expansions.

From the results of \cite{km} we expect that the
reduced dilaton field $\phi$, a singlet under the
one-loop duality, remains inert and does not pick
up any $O(\alpha')$ corrections. This comes about
because of the scheme we have adopted for the two-loop
terms (\ref{twola}). There the reduced
dilaton couples to the noninvariant terms exactly the
same as to the one-loop terms, implying that
the other degrees of freedom must be responsible for
restoring the symmetry. This implies 
that the duality transformation
on the dilaton is just a unity, and it remains decoupled
from the corrections. However, in a different
scheme, there would generally be corrections for the shifted
dilaton as well, coming from the fact that the dilaton
would interact with the other degrees of freedom via 
derivative couplings.
Moreover, we also expect that the modulus will get corrected
by an amount proportional to $(\nabla \sigma)^2$, for
if we look at the backgrounds with 
$V_{\mu}=W_{\mu}= H_{\mu\nu\lambda}=0$,
our case reduces precisely to that studied in \cite{km}.

Now, as we have explained above, we will demonstrate that
the noninvariant two-loop contributions will be absorbed
as the $O(\alpha')$ corrections in the one-loop action,
yielding a theory manifestly invariant under the corrected
$T$-duality transformation. To obtain
the precise form of the corrections
to the duality map, it is instructive to 
adopt an active approach, whereby we shift all
the one-loop level fields by an amount proportional
to $\alpha'$, and adjusts the shifts such that
they cancel the noninvariant couplings in
(\ref{noninvtact}). This approach will enable us
not only to evaluate the corrections, but also to
check their consistency with the anomaly cancellation.
We will discuss this in more detail below.

We define the shifts of the
one-loop degrees of freedom as follows:
\be
\label{shifts}
\sigma \rightarrow 
\hat \sigma + \alpha' \delta \sigma ~~~~
V_\mu \rightarrow 
\hat V_\mu + \alpha' \delta V_\mu  ~~~~
W_\mu \rightarrow 
\hat W_\mu + \alpha' \delta W_\mu ~~~~
 H_{\mu\nu\lambda} \rightarrow 
\hat H_{\mu\nu\lambda} + \alpha' \delta H_{\mu\nu\lambda} 
\ee
Starting from the one-loop 
action (\ref{actl}), we can show that 
the shifts induce the $O(\alpha')$ correction 
\ba
\label{shiftact}
\delta \Gamma = 
\alpha' \int d^d x \sqrt{g}e^{-2\phi}&\Bigl\{&
-2\pa_{\mu}\si\pa^{\mu}(\delta \si) 
- \frac{e^{2\si}}{2}[\delta \si V_{\mu\nu}^2 
+ V^{\mu\nu} \delta V_{\mu\nu}] \nonumber \\
&&+ 
\frac{e^{-2\si}}{2}[\delta \si W_{\mu\nu}^2 
- W^{\mu\nu} \delta W_{\mu\nu}] 
- \frac{1}{6} H^{\mu\nu\lambda} 
\delta H_{\mu\nu\lambda}\Bigr\}
\ea
In this formula we have replaced the
shifted degrees of freedom (denoted by a hat in 
(\ref{shifts})) by the original one-loop ones,
in the spirit of the active interpretation of
symmetries.
Our aim now is to determine the 
shifts of the fields such that the
correction of the action (\ref{shiftact})
precisely cancels the two-loop non-invariant terms
(\ref{noninvtact}). Here however 
we should first address the 
torsion field shift in more 
detail. As we have mentioned in section 2.,
at the one-loop level
the reduced torsion is a 
duality-singlet, just like the reduced
dilaton. The dilaton, on one 
hand, does not acquire any $O(\alpha')$
corrections, while even a cursory 
glance at the two-loop terms in
(\ref{noninvtact}) discloses that 
the torsion field must pick up
$O(\alpha')$ corrections if 
$T$-duality is ever to be restored.
One may thus wonder just why 
it may be necessary to correct a
field which is only a 
passive spectator in the arena of
$T$-duality, in order to 
assist the duality itself. This apparent 
dichotomy has been indicated at the end of the
section 2. It is resolved when 
we remember that the reduced torsion
couples to the gauge fields 
via the mixed Chern-Simons term. 
Given that the vectors 
transform nontrivially under duality, it
is clear that their corrections 
must play a key role in the
restoration of the duality at 
the two-loop level. Their coupling
to the torsion via the Chern-Simons 
terms however also induces 
the corrections of the torsion 
field. Thus the torsion 
assumes the role of a custodian 
field, with its corrections generated 
by gauge symmetries and needed 
to restore duality, while at the
same time preserving the form 
of the anomalous contributions. 
Indeed, as we will now show, 
gauge symmetry is very 
important in determining 
the form of the torsion corrections.

In section 2., we have explained that 
the components $B_{\mu\nu}$ 
by themselves do not specify a 
gauge invariant tensor. Rather, 
we must define the reduced torsion
potential according to 
$B_{\mu\nu} = \bar B_{\mu\nu} - W_{[\mu}V_{\nu]}$
(\ref{fintor}). The reduced field 
strength $ H_{\mu\nu\lambda}
= \nabla_{\mu} B_{\nu\lambda}
- \frac12 (W_{\mu\nu} V_\lambda + 
V_{\mu\nu} W_\lambda) + cyclic ~permutations$ 
is then fully compliant with the 
requirement of gauge invariance.
The presence of the torsion-gauge 
field couplings introduces $O(\alpha')$ 
corrections in the torsion, despite the fact that it
does not change under the lowest-order $T$-duality. 
The corrections are necessary because 
the only way to absorb
some of the terms quadratic in 
the torsion in (\ref{noninvtact}), for example,
is to include them in the 
one-loop torsion kinetic term.
Given that the corrections 
are induced by the torsion-gauge field 
couplings in the lowest 
order, they cannot be independent.
We have in fact anticipated this, 
for the following reason.
If the corrections to the 
$H$ field were independent and not
merely a consequence of the 
anomaly combination, we 
should expect that in higher loop 
orders the dilaton and 
the metric also start getting corrected 
in order to preserve duality - 
because they couple to the $H$ field,
they should also develop 
independent corrections. On the other hand, 
if we refer to the world-sheet 
description of the theory, we
can immediately see that the dilaton 
and curvature do not acquire gauge-field-dependent 
corrections like the torsion, because they are gauge-neutral
and anomaly-free.
Hence they should remain 
singlets under duality to all orders
in perturbation theory, and 
remain unchanged (modulo string field redefinitions) 
- an apparent contradiction. 
The only way to resolve 
this discrepancy
is to adopt the above 
viewpoint and express 
$\delta H_{\mu\nu\lambda}$ 
in terms of $\delta V_{\mu}$ and
$\delta W_{\mu}$, while 
maintaining gauge invariance.
How are we to accomplish this? 
First, assuming that 
$\delta H$ originates only 
from $\delta V$ and $\delta W$,
and imposing that gauge 
invariance must be respected by 
the corrections (meaning 
that the corrections must preserve
the functional form of the 
definition of $\hat H$ and the
associated Bianchi identity,
because the anomaly arises 
only in the lowest order 
of the theory and does not receive
other nontrivial corrections) 
we find that to linear order,
\be
\label{delh}
\delta H_{\mu\nu\lambda} = 
\nabla_{\mu} \delta B_{\nu\lambda} - \frac12
(W_{\mu\nu} \delta V_\lambda 
+ V_{\mu\nu} \delta W_{\lambda})
- \frac12
(\delta W_{\mu\nu} V_\lambda 
+ \delta V_{\mu\nu} W_{\lambda})                
\ee
Since the starting one-loop 
action is gauge invariant, 
and the corrections are 
generated by higher-order terms in 
perturbation theory, they can
only depend on gauge-invariant variables.
Therefore, $\delta \sigma$, 
$\delta V_{\mu}$ and $\delta W_{\mu}$
must be themselves gauge 
invariant. However, in (\ref{delh})
we find terms which 
appear to depend explicitly on the 
one-loop gauge potentials 
$V_{\mu}$ and $W_{\mu}$, which are
not gauge-invariant! Thus, to restore 
gauge invariance of $\delta H_{\mu\nu\lambda}$,
$\delta B_{\mu\nu}$ must also 
involve corrections, which will
cancel exactly the explicit 
gauge-dependent terms in $\delta H$. Indeed, 
we see that there are no such 
gauge-dependent terms anywhere in 
the two-loop part of the action.
So, 
from $B_{\mu\nu} = \bar B_{\mu\nu} - W_{[\mu} V_{\nu]}$,
we see that
\be
\label{hatbcor1}
\delta B_{\mu\nu} = \delta \bar B_{\mu\nu} 
- \delta W_{[\mu} V_{\nu]} - W_{[\mu} \delta V_{\nu]}
\ee
If the gauge transformations of the Kaluza-Klein
gauge field $V_\mu \rightarrow 
V_{\mu} + \pa_{\mu} \omega$ 
are to remain correct in 
all orders of perturbation
theory, which they must for 
they only correspond to special 
diffeomorphisms, and no string theory
has anomalies which can break these symmetries 
quantum-mechanically, then the gauge transformation
rules for all fields must
remain the same as they are in the 
classical limit. The corrections must 
assemble themselves in such a 
way that the gauge symmetry is preserved
beyond the classical limit.
If we apply this to the corrected fields 
$\hat B_{\mu\nu} = \bar B_{\mu\nu} 
- \delta \bar B_{\mu\nu}$    
and $\hat W_\mu = W_\mu 
- \delta W_{\mu}$, we get
\be
\label{torgt1}
\hat B'_{\mu\nu} =  \hat B_{\mu\nu} 
+ \hat W_{\mu} \pa_\nu \omega 
- \hat W_{\nu} \pa_\mu \omega 
\ee
Separating these fields 
as the classical parts plus the
correction, and using the 
gauge transformation properties
of the classical terms,
we find that the correction must satisfy 
an analogous transformation law:
\be
\label{corrtorgt}
\delta \bar B'_{\mu\nu} =  \delta \bar B_{\mu\nu} 
+ \delta W_{\mu} \pa_\nu \omega 
- \delta W_{\nu} \pa_\mu \omega 
\ee
Since $\omega$ is just the 
classical transformation, and the corrections
$\delta \bar B_{\mu\nu}$ come 
from the $V$ and $W$ gauge
fields, we can see immediately 
that in order for the last equation
to be true we must have
\be
\label{gauginvcorr}
\delta \bar B_{\mu\nu} = b_{\mu\nu} 
+ \delta W_{\mu} V_{\nu}
- \delta W_{\nu} V_{\mu}
\ee
The $\delta W$-dependent 
corrections are dictated by gauge
symmetry, and $b_{\mu\nu}$ is 
an additional correction to $B$, which
must be gauge invariant, and 
is needed to cancel the 
terms in (\ref{noninvtact}) 
that depend on derivatives of
$H$. Because we are working 
with the action to order $\alpha'$,
by dimensional arguments 
these terms must be quadratic in
derivatives of $V$ and $W$. 
This determines $b_{\mu\nu}$
up to a constant: $b_{\mu\nu} = c 
W_{\lambda[\mu} V^{\lambda}{}_{\nu]}$.
Substituting it in 
the expression for $\delta B$,
we find that
\be
\label{gauginvthatb}
\delta B_{\mu\nu} = 
c W_{\lambda[\mu} V^{\lambda}{}_{\nu]}
+ \delta W_{[\mu} V_{\nu]} 
+ \delta V_{[\mu} W_{\nu]}
\ee
Now we can substitute this 
expression back into the formula for 
$\delta H$ (\ref{delh}). 
Rewriting that equation as
$\delta H_{\mu\nu\lambda} = 
3 \nabla_{[\mu} \delta B_{\nu\lambda]}
- (3/2) V_{[\mu\nu} \delta W_{\lambda]} 
- (3/2) W_{[\mu\nu} \delta V_{\lambda]}
- (3/2) \delta V_{[\mu\nu} W_{\lambda]} 
- (3/2) \delta W_{[\mu\nu} V_{\lambda]}$,
after some simple algebra we find 
that the gauge-dependent terms cancel out:
\be
\label{gauginv3tor}
\delta H_{\mu\nu\lambda} = 
3c \nabla_{[\mu} \bigl(W_{\nu}{}^{\rho}
V_{\lambda]\rho} \bigr) 
- 3 V_{[\mu\nu} \delta W_{\lambda]} -
3 W_{[\mu\nu} \delta V_{\lambda]}
\ee
Having thus determined $\delta H$, all
we need to do now is insert 
it in the action, and determine
$\delta \sigma$, $\delta V$ and 
$\delta W$ which cancel (\ref{noninvtact})
and restore $T$-duality.

The unique explicit form of the 
corrections which lead to the cancellation
between (\ref{noninvtact}) and (\ref{shiftact}) is
\ba
\label{corrs}
\delta \si &=& -2 \lambda_0 \dstwo
-\frac{\lambda_0}4\ets Z
-\frac{\lambda_0}4\emts T\nonumber\\
\delta V_{\alpha} &=&
2\lambda_0 V_{\alpha\rho}\pa^{\rho}\si
-\frac{\lambda_0}2 H_{\alpha\beta\gamma}
W^{\beta\gamma}\emts\nonumber\\
\delta W_{\alpha} &=&
2\lambda_0 W_{\alpha\rho}\pa^{\rho}\si
+\frac{\lambda_0}2 H_{\alpha\beta\gamma}
V^{\beta\gamma}\ets
\ea
with the constant $c=-2 \lambda_0$.
In order to derive this, all one needs to do
is simply collect the like terms in the
sum of (\ref{noninvtact}) and (\ref{shiftact}), demand
that they vanish, and find the set of linear
equations for the free parameters, (which have
a nonzero determinant), leading
to the solution (\ref{corrs}). The resulting
two terms in the action, the renormalized one-loop
part and the invariant two-loop part, are then 
manifestly invariant under the one-loop level
form of the $T$-duality map $\sigma \leftrightarrow
-\sigma$, $V_{\mu} \leftrightarrow W_{\mu}$, acting
on the corrected fields. If we return then to the 
original one-loop level degrees of freedom (unhatted ones
in the equation (\ref{shifts})), we 
can interpret the $O(\alpha')$ shifts as the
two-loop corrections to the action of the $T$-duality map
on these degrees of freedom. Using this, 
we can rewrite the two-loop $T$-duality
transformation equations as
follows:
\ba
\label{corrTd}
\sigma &\rightarrow & -\sigma - 4 \aplz 
(\nabla \sigma)^2 - \frac{\aplz}{2} [\ets Z + \emts T]
\nonumber \\
V_{\mu} &\rightarrow & W_{\mu} -4 \aplz W_{\mu\nu} \nabla^{\nu} \sigma -
\aplz  H_{\mu\nu\lambda}V^{\nu\lambda} \ets \nonumber \\
W_{\mu} &\rightarrow & V_{\mu} -4 \aplz V_{\mu\nu} \nabla^{\nu} \sigma
+ \aplz  H_{\mu\nu\lambda} W^{\nu\lambda} \emts \\
H_{\mu\nu\lambda} &\rightarrow&  H_{\mu\nu\lambda}
-12 \aplz \nabla_{[\mu} \bigl(W_{\nu}{}^{\rho} V_{\lambda]\rho} \bigr)
- 12 \aplz V_{[\mu\nu}W_{\lambda]\rho}\nabla^{\rho} \sigma \nonumber \\
&&- 12 \aplz W_{[\mu\nu}V_{\lambda]\rho}\nabla^{\rho} \sigma 
- 3\aplz \bigl(\ets V^{\rho\sigma} V_{[\mu\nu}
- \emts W^{\rho\sigma} W_{[\mu\nu} \bigr)  H_{\lambda]\rho\sigma} 
\nonumber
\ea
Then, the full reduced action, containing all one-
and two-loop contributions (\ref{actl}), (\ref{noninvtact}) and
(\ref{invact}) is invariant under (\ref{corrTd}) to order
$\alpha'$, as one can check by directly applying these
transformation rules. This is our final result.

Before closing this section, we note in passing
that the $\beta$-function transformations under
duality to two loops can be obtained as an easy 
byproduct of our approach. Since with the help
of the corrections (\ref{shifts}) 
and (\ref{corrs}) we can recast
the action with the two-loop terms in the form that is
invariant under the one-loop duality map, it
should be clear that the functional derivatives of 
this renormalized action transform under duality
in the same way as the functional derivatives of the
one-loop action. To find explicitly 
how the functional derivatives transform
at the two-loop level, then, all we need to do is 
to write down the one-loop equations of motion, 
obtained from varying the action(\ref{actl}), and 
simply replace the original one-loop fields
by the corrected fields (\ref{shifts}). Then, if we expand the
corrected fields in $\alpha'$, collect the $O(\alpha')$
contributions, and reverse their overall sign, we get the 
$O(\alpha')$ corrections to the transformation 
rules of the functional derivatives. 
The next step then is to resort to the local linear
relationship between the functional derivatives of
(\ref{twola}) with the $\beta$-functions, which we have 
discussed at the beginning of the section 2.
Inverting this relationship, and correctly accounting for
the transformation properties of this linear map,
we would obtain the two-loop form of the transformation
properties of the $\beta$-functions under $T$-duality -
at and away from the conformal points.
Presenting the explicit form of these
rules is beyond our goal here, and we defer it to
the future.

\section{Conclusion}
\setcounter{equation}{0}

In this paper we have examined the
$O(\alpha')$ two-loop corrections in string theory and their 
effect on, and compliance with, $T$-duality. Our approach
was based on the effective field theory of the 
model-independent zero mass sector, consisting of the
graviton, dilaton and torsion. The two-loop
terms are obtained from the string amplitude calculations,
and put in the manifestly unitary form with the string 
field redefinitions. This has in fact allowed us to 
ascertain that our calculations apply to different
string theories (heterotic, bosonic and superstring (in which case
$T$-duality of course is trivially extended to two loops, because
the corrections vanish identically)), 
parametrized by a single constant. Adopting
a specific string subtraction scheme and the form
of the two-loop effective action this scheme dictates,
we have focused on the string backgrounds with a
single isometry, and have shown that the theory
is invariant under two-loop corrected $T$-duality. 
We have arrived at the form of the corrections 
by an iterative reformulation of the $\alpha'$
expansion: any $O(\alpha')$ terms found to violate
the one-loop form of duality were interpreted to induce
$O(\alpha')$ corrections in the original duality map. 
The explicit form of duality transformations we have
found contains terms linear in $\alpha'$, which should be
thought of as the first subleading terms of the Taylor
expansion of the duality map in $\alpha'$. Our
computations thus are in full agreement with the
expectation that $T$-duality is perturbatively 
exact. One unusual feature of our approach 
is that solutions of the lowest-order
action which have unbroken supersymmetries, and hence
are BPS states, do not retain their form when the two-loop
terms are included. In other words, the two-loop 
contributions do not cancel among themselves
on supersymmetric backgrounds. As a result, when
our duality corrections are applied to BPS states,
they contain nonvanishing terms to $O(\alpha')$.
While this may sound odd, given the current 
lore \cite{exact}, one should remember 
that while for BPS states there exists a scheme in which 
the classical solutions are exact 
to all orders in the $\alpha'$ expansion, this of course
need not be true in any scheme. In fact, 
those terms among the two-loop
corrections which we consider and 
which do not vanish on BPS backgrounds
should be removed by string field redefinitions.
As an example, one can take a special class of
BPS states, given by the four-dimensional
extremal Reissner-Nordstrom black hole metric 
and with two purely electric 
gauge fields $V_{\mu}=W_{\mu} = q/r \delta_{\mu0}$, and
two purely magnetic ones, 
$v_{\mu}=w_{\mu} = q \cos(\theta)\delta_{\mu\phi}$.
For this solution, $\phi=\sigma=H_{\mu\nu\lambda}=0$.
On the other hand, this solution can be ``oxidized"
to six dimensions, where the gauge fields $V_{\mu}$,
$v_{\mu}$ are returned to the metric, and $W_{\mu}$, 
$w_{\mu}$ to the torsion.
Since the field redefinitions to order $\alpha'$ depend 
only on the scalars and rank-two tensors of the solution
which are quadratic in derivatives, with little algebra
and the help of the equations of motion, 
one can verify that the only nontrivial redefinition
on this background is $\delta \bar g_{\mu\nu} = \alpha' \lambda_0
\kappa \bar R_{\mu\nu}$. The parameter $\kappa$ 
can then be adjusted to remove the correction of 
the metric component $\bar g_{yy}$, or $\si$.
The remaining terms should then remove the corrections
for the metric and gauge fields of the reduced theory.
We will not delve on the details here. Suffice it to say that, 
in some sense, these terms behave like
gauge degrees of freedom. Finally, we have 
indicated in closing how our 
approach can be utilized to determine
transformation properties of the string 
$\beta$-functions to two loops.

A rather remarkable feature of our calculation
has been its reliance on gauge invariance of the
reduced action, where the gauge sector consists
of the Kaluza-Klein gauge fields which arise from 
dimensional reduction of the metric and torsion. 
The anomalous transformation
properties of the torsion, reflected in the 
Chern-Simons terms in the definition of the 
torsion field strength play a crucial role in 
restoring two-loop duality, while at the
same time maintaining gauge invariance and the
lowest-order form of the anomaly. This perhaps
should not come as a complete surprise. As has been
pointed out by Maharana and Schwarz, who discovered
the Chern-Simons terms in the reduced theory \cite{mahsch}, 
the anomaly was essential in rendering the one-loop theory
$T$-duality invariant. This role of the anomaly seems to
persist to two loops, and raises an interesting possibility
that the concepts of the anomaly and $T$-duality 
may somehow be related in the full quantum theory beyond
the effective action limit (M-theory). 
An investigation of any such presumed relationship,
however, may demand resorting to 
a different, nonperturbative, approach, due to 
the complexity of the higher-loop counterterms. 

\vspace{1cm}
{\bf Acknowledgements}

K.A.M is grateful to Gabriele Veneziano for
earlier collaboration 
and for discussions on $O(d,d)$ symmetry.
N.K. would like to thank the Theory Division at
CERN, where this work has 
begun, and in particular G. Veneziano for kind 
hospitality. We are also indebted to 
R.R. Khuri and R.C. Myers for helpful conversations 
and comments on the manuscript.
This work was supported 
in part by NSERC of Canada.

\end{document}